\documentstyle[epsfig]{l-aa}

\begin{document}

\def\ltsima{$\; \buildrel < \over \sim \;$}
\def\simlt{\lower.5ex\hbox{\ltsima}}

   \thesaurus{08.02.1 -- 08.14.2 -- 13.25.5 -- 08.09.2: Am Herculis}
   \title{BeppoSAX observations of AM Herculis in intermediate and high states}

   \author{G. Matt 
          \inst{1}
    \and D. de Martino
          \inst{2}
    \and B.T. G\"ansicke
          \inst{3}
    \and I. Negueruela 
          \inst{4}
    \and R. Silvotti 
          \inst{2}
    \and J.M. Bonnet--Bidaud 
	 \inst{5}
    \and M. Mouchet
          \inst{6,7}
    \and K. Mukai
          \inst{8}
          }

   \offprints{G. Matt, matt@fis.uniroma3.it}

   \institute{Dipartimento di Fisica, Universit\`a degli Studi ``Roma Tre", 
              Via della Vasca Navale 84, I--00146 Roma, Italy
    \and Osservatorio Astronomico di Capodimonte, Via Moiariello 16, 
         	I-80131 Napoli, Italy
   \and Universit\"atssternwarte G\"ottingen, Geismarlandstrasse 11, D-37083
                 G\"ottingen, Germany
    \and SAX/SDC Nuova Telespazio, Via Corcolle 19,  I--00131 Roma, Italy
   \and CEA, DSN/DAPNIA/Service d'Astrophysique, CEN Saclay, F-91191 Gif sur
         Yvette Cedex, France
    \and DAEC, Observatoire de Paris, Section de Meudon, F-92195 Meudon Cedex, France
    \and Universit\'e Denis Diderot, 2 Place Jussieu, F-75005 Paris, France
    \and NASA/GSFC, Code 668, Greenbelt MD 20771, USA
   }

   \date{Received 3 January 2000 / Accepted }

   \maketitle

   \begin{abstract}
Temporal and spectral analyses from BeppoSAX observations of AM Her performed
during both an intermediate and a high state are presented and discussed.
Optical observations taken a few days after the X--ray ones are also presented.

During the intermediate state observation, the source was in its ``normal",
one--pole accretion mode. In the high state 
it switched to an hitherto unobserved atypical ''two-pole''
accretion mode, with significant soft and hard X-ray emission from
both poles. The emission from the second pole is much softer than that
from the primary pole, while the soft X-ray excess of the primary pole
is fairly modest in this accretion mode. These facts
suggest that accretion onto the secondary
is mainly due to blobs penetrating deeply in the photosphere, 
while that on the primary pole is mostly via a more homogeneous column, 
giving rise to the classical standing shock. 
A strong X-ray  flaring activity is also  observed
in the soft X-ray band, but not the hard X-ray and optical emissions 
indicating that flares are due to inhomogeneous blobby-accretion events.
   \end{abstract}

%

\section{Introduction}

 Polars, a subgroup of magnetic Cataclysmic Variables (mCVs),
 contain a highly magnetized white dwarf with polar field strengths
ranging from  $\sim$ 10
MG $\sim$ to 230 MG (see Beuermann 1998 and references therein), and which
 accretes from a late--type main sequence star. The
magnetic field of the white dwarf is strong enough to phase--lock its
rotation with the orbital period.
These systems are strong X-ray emitters in both soft and hard X-ray bands
 (see review by Cropper 1990).  While hard X-rays and optical
cyclotron radiation are emitted from
a standing shock above the white dwarf surface, 
soft X-rays originate from hot photospheric regions heated
either by irradiation from the post-shock plasma (Lamb \& Masters 1979) or by
dense plasma blobs carrying their kinetic energy deep into the atmosphere
(Kuipers \& Pringle 1982). 
Irradiation is important primarily for flow rates sufficiently low for
the shock to stand high above the surface. However, a large fraction
of the reprocessed radiation appears in the UV rather than at soft
X-rays, as shown quantitatively in the prototypal system AM\,Her
(G\"ansicke et al. 1995). 

While the soft X--ray component is in general adequately fitted by a
black--body spectrum with a temperature of a few tens of eV (even if more
sophisticated models are sometimes needed: see e.g. Van Teeseling et al. 1994), 
it has recently become clear that a simple thermal plasma
model is no longer adequate in reproducing the hard X--ray emission of Polars. 
Reflection
from the white dwarf surface, complex absorption and multi--temperature
emission may contribute significantly to the spectrum above a few tenths of
keV (e.g. Cropper et al. 1998 and references therein). 

Polars are known to display long--term luminosity variations on several
timescales ranging from tens of days to years. These variations
are thought to be related to changes
 in the mass transfer rate from the secondary star. The best studied example
 is again the brightest polar AM\,Her (Cropper 1990; G\"ansicke et al. 1995).
 Such changes come along with spectral variations in all energy bands,
 indicating changes not only in the spectral parameters of the primary
 radiation but also in the reprocessed emission. The relative proportion of
 hard and soft X-ray emissions is expected to be sensitive to the actual
 accretion luminosity. This long--term variability is scarcely monitored
 and important questions related to changes in accretion parameters and in
accretion modes (such as switching from one--pole to two--pole accretion),
 as well as to the processes of energy release, are still unanswered.

To properly address some of these issues, a  long--term monitoring program 
of the
 prototype AM\,Her was set  up with the BeppoSAX satellite in
 order to investigate its long--term behaviour on a wide 
(0.1--100 keV) X-ray energy band
 simultaneously. These observations have been complemented with optical
 photometry for a more comprehensive study.


\section{Observations and data reduction}

\subsection{The BeppoSAX data} 

\begin{figure}
\epsfig{file=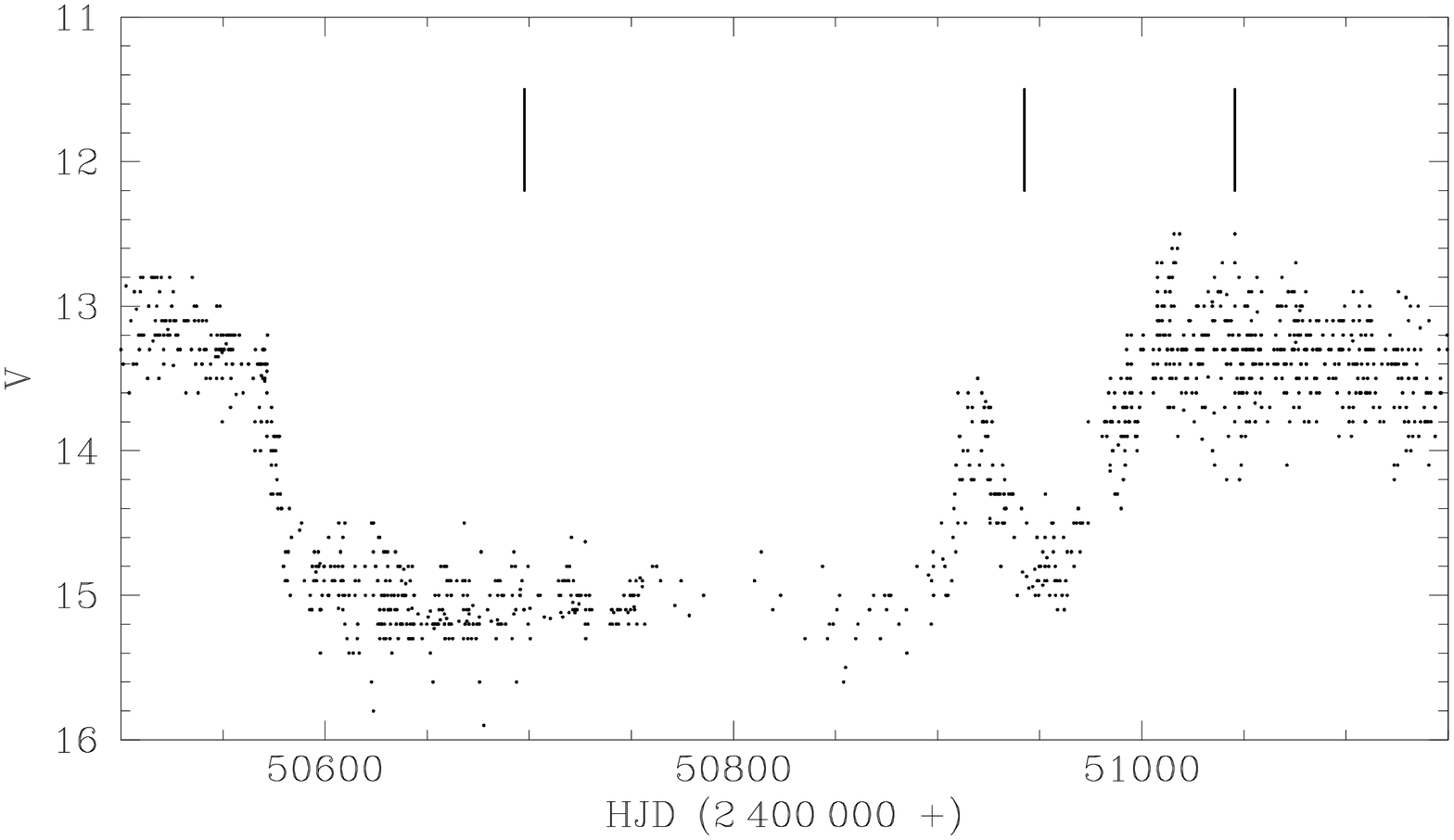, height=6.cm, width=8.5cm}
\caption{ The secular optical light curve obtained from AAVSO and VSNET
          showing the three pointings of Sept. 1997, May 1998, and
          August 1998, corresponding to epochs when AM\,Her was in a deep
          low state, an intermediate and high states respectively.} 
\label{aavso}
\end{figure}

BeppoSAX (Boella et al. 1997) observed AM Her
three times, with the source in very
different states: a deep low state, an intermediate state, a high state.
The journal of the observations is in Table~\ref{journal}. The epochs 
of the three BeppoSAX observations are also indicated in Fig.~\ref{aavso},
superimposed on the secular AAVSO and VSNET optical light curves.

\begin{table}
\caption{ Log of BeppoSAX and Optical observations.}
\label{journal}
\begin{tabular}{|c|c|c|}
\hline
~ & ~ & ~ \cr
Date & MECS Exp. & Flux$^{1}$/MECS~count~rates$^{2}$\cr
~ & time~(s) & \cr
~ & ~ & ~ \cr
\hline
~ & ~ & ~ \cr
Sep. 6, 1997 & 24700 & 0.085/0.013 \cr
 May 8, 1998 & 33500 & 1.8/0.22    \cr
Aug. 12, 1998 & 80600 & 12/1.35 \cr
~ & ~ & ~ \cr
\hline
~ & ~ & ~ \cr
Date & Filter & Exp. time  \cr
~ & ~ & (s) \cr
\hline
~ & ~ & \cr
May 20, 1998 & No filter & 6120 \cr
May 21, 1998 & No filter & 12600 \cr
Aug. 20, 1998 & V & 13320 \cr
Aug. 21, 1998 & V & 10440 \cr
Aug. 22, 1998 & B & 14040 \cr
Aug. 23, 1998 & U & 11880 \cr
~ & ~ & ~ \cr
\hline
\end{tabular}
\medskip
~\par
$^{1}$ 2--10 keV, phase averaged flux in units of 10$^{-11}$ erg cm$^{-2}$ s$^{-1}$;
see de Martino et al (1998) and the text for best fit models. \par
$^{2}$ In units of cts  s$^{-1}$.

\end{table}

\subsubsection{ Low state}

The source was observed in September 1997
in the midst of a prolonged low state (Fig.~\ref{aavso}). During the
first four hours the flux level was comparable to that normally observed
in low states, when the X--ray emission is still dominated by accretion onto
the white dwarf. Then, with a sudden drop by a factor of about 7 in 40 min,
the source went into the deepest low state ever observed (see Table\,1), in which the X--ray
emission was probably at least in part due to coronal emission from the
secondary star. Such a  fast change in the accretion rate
is not easy to accommodate with current scenarios of mass transfer from
the secondary and it represents a serious challenge to theoretical models.

The results of this observation have been already published 
(de Martino et al. 1998), and will not discussed here
any further.

\subsubsection{Intermediate state}

The second BeppoSAX observation of AM Her was performed in May 1998, after
a brief episode of relatively high flux (Fig.~\ref{aavso}).
In X--rays, the source was at an intermediate flux level, but high enough to
permit the temporal and spectral analysis with three BeppoSAX Narrow Field
Instruments: LECS (0.1-10 keV), MECS (1.8-10.5 keV) and PDS (15-200 keV).
LECS spectra and light curves have been extracted within a 8' radius circular
region centred on the source. 
While for temporal analyses the full LECS band can be used, the
spectral analyses have been restricted below 4 keV due to remaining
calibration problems at higher energies. The adopted extraction region
for the MECS is instead 4'. For both instruments, we have subtracted the
background measured in the same detector regions during 
blank field observations. 
In both instruments the source flux is in any case much larger than 
the background over the whole energy
band. For the PDS, which is a collimated instrument, the background 
is monitored by continuously switching the detectors on and
off the source, with a rocking time of 96 seconds, 
in such a way that 2 out of 4 detectors are always on source. In the
following, PDS exposure times refer to the total time during which the 
source is observed (whatever the pair actually on source), while count rates
refer to all 4 detectors. 

Net exposure times and time averaged background subtracted
count rates are respectively:  16749 s, 0.1273$\pm$0.0029 (LECS); 33556 s, 0.2196$\pm$0.0026 (MECS, 2
detectors); 30596 s, 0.114$\pm$0.068 (PDS). The shorter LECS exposure 
is due to operational limits on the LECS detector, which 
 can be operated only during satellite nights.

\subsubsection{High state}

In August 1998, when BeppoSAX observed AM\,Her for the third 
time, the source had already recovered its optical high
state (Fig.~\ref{aavso}). The X--ray flux was also at high level.

Data reduction is as described in the previous subsection.
Net exposure times and background subtracted time averaged count
rates are respectively: 40053 s, 0.9200$\pm$0.0048 (LECS);
80631 s, 1.3510$\pm$0.0041 (MECS,  2 detectors);
73214 s, 1.017$\pm$0.034 (PDS).

\subsection{Optical photometry}

AM\,Her was observed from the Loiano 1.5\,m telescope with a 2-channel
photoelectric photometer in  1998 May 20 and 21 and August 20, 21, 22
and 23 a few days after the SAX pointed observations. 
It was at V$\sim$ 14.5\,mag during May run and white light photometry was
acquired with an integration time of 5\,s
for 1.7\,h on May 20 and 3.5\,h on May 21. In August it was at 
V=13.51$\pm 0.15$\,mag and B=13.65$\pm 0.15$\,mag. During each night
AM\,Her was observed with an integration time of 1\,sec 
for 3.7\,h  and for 2.9\,h 
 in  the V filter on August 20 and 21, respectively; in the band B on August 
22 for 3.9\,h and in U filter on August 23 for 3.3\,h. 
At irregular intervals of typically 60--120\,min sky measures were performed
in both channels for about 1 min.
Photometric data have been reduced taking into account sky subtraction and 
differential extinction.
Moreover, the V and B data were calibrated using a set of Landolt standards.

\section{Temporal analysis}

\subsection{The X-ray Intermediate State}

\begin{figure}
\epsfig{file=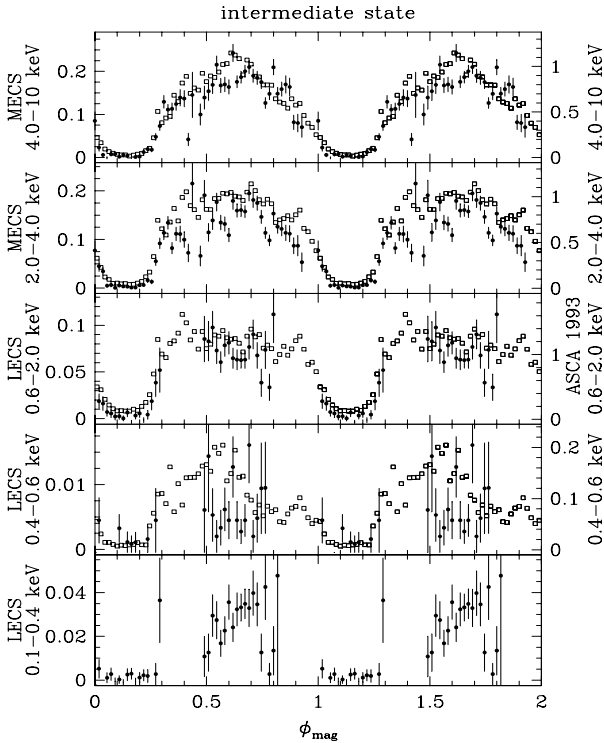, height=10.cm, width=8.cm}
\caption{ LECS (0.1-0.4\,keV),(0.4-0.6\,keV), (0.6-2.0\,keV) and MECS
           (2-4\,keV), (4-10\,keV) count rate light curves (black dots) 
           during the intermediate state of May 1998 folded  in 55 orbital
           bins. The intermediate/high state light curves observed by ASCA 
           (open squares) in 1993 are also reported for comparison.
Both the BeppoSAX data and the ASCA
data are scaled individually to their respective maximum count rates.
           At this epoch AM\,Her recovered its ``normal" one-pole accretion
           mode. Note the absence of significant flux during orbital minimum,
           corresponding to the self-occultation of the primary pole.} 
\label{interm_lc}
\end{figure}

Both LECS and MECS detectors reveal the typical orbital modulation
 at $\rm P_{orb} = 3.09\,hr$. Adopting the linear  polarization ephemeris
quoted in Heise \& Verbunt (1988), the folded light curves in different 
energy bands are shown 
in Fig.~\ref{interm_lc}. These are single peaked with a broad maximum
(bright phase) centered at  $\Phi_{\rm mag} = 0.6$ and extending 
between $\phi$=0.4--0.8 and with count rates dropping 
to zero between  $\Phi_{\rm  mag} =0.02 - 0.22$ (faint phase). 
The general shape of the soft and hard X-ray light
curves is typical for the intermediate/high state of AM\,Her, and the 
BeppoSAX light curves resemble closely those obtained with ASCA 
in 1993 (Ishida et al. 1997) 
during a prolonged high state ($V\approx13.4$). The ASCA light
curves are included in Fig.~\ref{interm_lc} for comparison. 

 Hence during this epoch AM\,Her was in its ``normal" accretion state, 
dominated
 by the primary pole, which is self-occulted during minimum.
As in the ASCA or GINGA observations, no significant
emission from the secondary pole is observed. Noteworthy 
is the similarity between the
 high energy ($\geq$ 0.6\,keV) light curve shapes of BeppoSAX and ASCA.
 An energy dependence of the light curves is also apparent; the curves are
more sinusoidal
 at higher energies (above 4\,keV) while more structured at lower energies.
Unlike the faint phase count rates,
 which behave similarly to those at higher energies (i.e. reaching the zero
 level), the bright phase in the softest range (0.1-0.4\,keV) increases between
 $\rm \Phi_{orb} \sim $ 0.5--0.7, possibly suggesting a secondary minimum,
not covered by the data, at earlier phases. Indeed, the shape of this 
low-energy LECS light curve departs from the soft X-ray ASCA 
light curve (0.4--0.6 keV) which in turns compares well with the EUVE light curve of 
AM Her obtained during a high state, revealing 
revealing geometric absorption effects in the emission of the main
pole (Paerels et al. 1996, Sirk \& Howell 1998).

 Furthermore, a tendency in the hardness ratios to soften at the rotational
 minimum is inferred,  [(4-10\,keV) - (1-4\,keV)]/[(1-4\,keV)] + (4-10\,keV)]
 being -0.29 while it is  -0.08 during the bright phase (see also Sect. 3.2 
 and 4).  Also,
 during the bright phase a significant flickering activity of $\sim36\%$ with
 respect to the average is observed in the MECS detector on timescales 
ranging between 2\,mins  and 15\,mins, while the poor statistics in the LECS
detector prevent any detection of rapid fluctuations.



\begin{table*}
\centering
\caption{ Best fit parameters for the intermediate state, bright phase} 

\label{fit_is}
\vspace{0.05in}
\begin{tabular}{lccccccccc}
\hline
\hline
~ & ~ & ~ & ~ & ~ & ~ & ~ & ~ & ~  \cr
\# & N$_{\rm H}$$^a$ &  C$^b$ & $kT_{\rm bb}$ & $kT_{\rm h}$$^c$  
& $A_Z$$^d$  & E.W.$^e$  & $R^f$ &  $\chi^2$/d.o.f. ($\chi^2_r$)   \cr
~ & (10$^{22}$ cm$^{-2}$) & ~ & (eV) & (keV) & ~ & (eV) & ~ & ~ \cr
~ & ~ & ~ & ~ & ~ & ~ & ~ & ~ & ~  \cr
\noalign {\hrule}
~ & ~ & ~ & ~ & ~ & ~ & ~ & ~ & ~  \cr
1 & ~ & ~ & 19.3$^{+3.0}_{-1.4}$ & 23.1$^{+7.4}_{-4.2}$ & 
2.77$^{+1.10}_{-0.65}$ & ~ & ~ & 174/127   (1.37) \cr
2 & 2.1$^{+1.6}_{-1.1}$ & 0.40$^{+0.09}_{-0.09}$ & 19.6$^{+2.7}_{-1.5}$  
& 13.8$^{+4.0}_{-2.6}$ & 1.51$^{+0.60}_{-0.41}$  & ~ & ~ & 143/125 (1.15) \cr
3 & 1.35$^{+1.45}_{-0.91}$ & 0.37$^{+0.11}_{-0.11}$ & 19.8$^{+2.5}_{-1.4}$ 
& 14.4$^{+7.1}_{-5.7}$ & 
1.24$^{+0.79}_{-0.65}$  & 150$^{+120}_{-147}$  & 0.71$^{+3.17}_{-0.71}$  & 135/123 (1.10) \cr
~ & ~ & ~ & ~ & ~ & ~ & ~ & ~ & ~  \cr
\hline
\hline
\end{tabular}
~\par
$^a$ Column density of the partial absorber.\par
$^b$ Covering fraction of the partial absorber.\par
$^c$ Plasma temperature.\par
$^d$ Metal abundance in units of the cosmic value (Anders \& Grevesse 1989).\par
$^e$ Equivalent width of the 6.4 keV fluorescent iron line.\par
$^f$ Relative normalization of the reflection component (see text).
\end{table*}

\subsection{The X-ray High State}

\begin{figure}
\epsfig{file=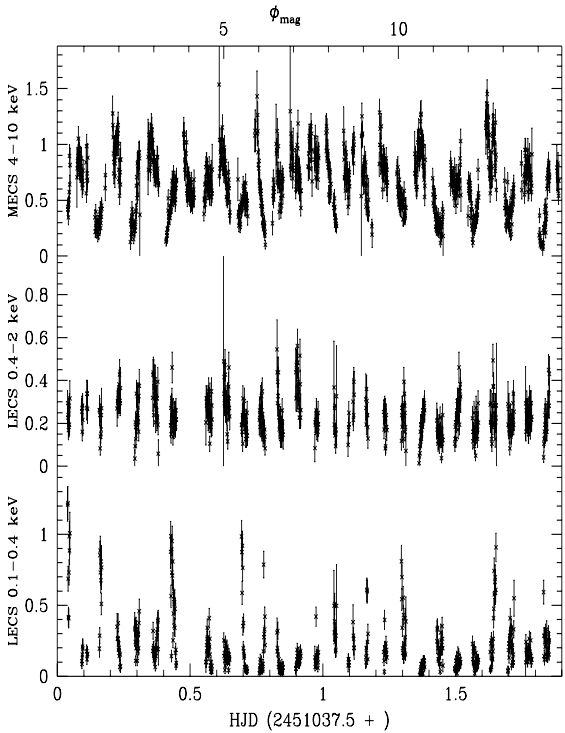, height=10.cm, width=8.cm}
\caption[] { LECS (0.1-0.4\,keV),(0.4-2\,keV) and MECS (4-10\,keV) 90\,sec
           binned light curves of AM\,Her during the high state in August
           1998 showing a soft flaring activity while the orbital 
modulation dominates in the hard X-rays.}
\label{high_unfold_lc}
\end{figure}

The August 1998 high state light curves, shown in  Fig.~\ref{high_unfold_lc}, 
are characterized by a strong  variability. While at energies below 
0.4\,keV an intense flaring activity is observed, at higher energies
the variations are periodic and associated with the white dwarf rotation.

\begin{figure}
\epsfig{file=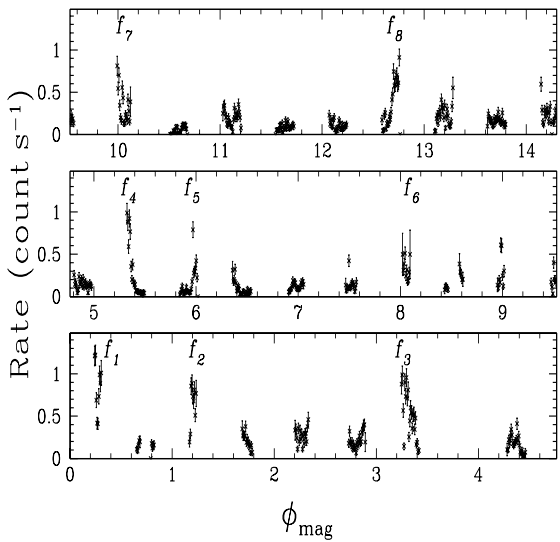, height=8.cm, width=8.cm, angle=-90}
\caption[] { The LECS (0.1-0.4\,keV) 90\,sec binned light curve showing 
strong flaring activity of the order of several tens of minutes.}
\label{soft_unfold_lc}
\end{figure}

At least 8 flares can be identified in the soft 0.1-0.4 light curve
(Fig~\ref{soft_unfold_lc}). The typical duration of these events is longer than
the target visibility during one BeppoSAX orbit, preventing the
complete coverage of any of the flares. Count rates during these 
events increase up to a factor of 6 from a mean 
``quiescence'' level of $\sim 0.17\, \rm cts\,s^{-1}$. The rise and decay of 
the better sampled events, as for instance the rise of flare {\it $f_{8}$}
and the decay of flare {\it $f_{4}$}
have been fitted with an exponential slope resulting in an $e$--folding time
$\tau$ of 8 and 22 min, respectively. 
These events are randomly detected at all magnetic phases and two of 
them, {\it $f_{6}$} and 
{\it $f_{7}$},  indicate that activity is not only related
to the main accreting pole.  While no hardness ratio can be
determined in the narrow 0.1-0.4 keV LECS band, these flares are clearly soft 
(see also Sect. 4). A soft X-ray 
flaring  behaviour during high states is not uncommon
for AM\,Her (Priedhorsky et al., 1987; Ramsay et al. 1996), but with much 
shorter time scales (from $\sim$ 30\,s to a few minutes). We note, however, 
that the statistics of LECS data does not allow us to exclude that the 
observed flares are constituted by train (or series) of shorter events.

In order to inspect the orbital modulation also in the softest channels
($\leq$ 0.4\,keV), the eight detected flares have been removed before folding 
count rates along the magnetic phases.

\begin{figure}
\epsfig{file=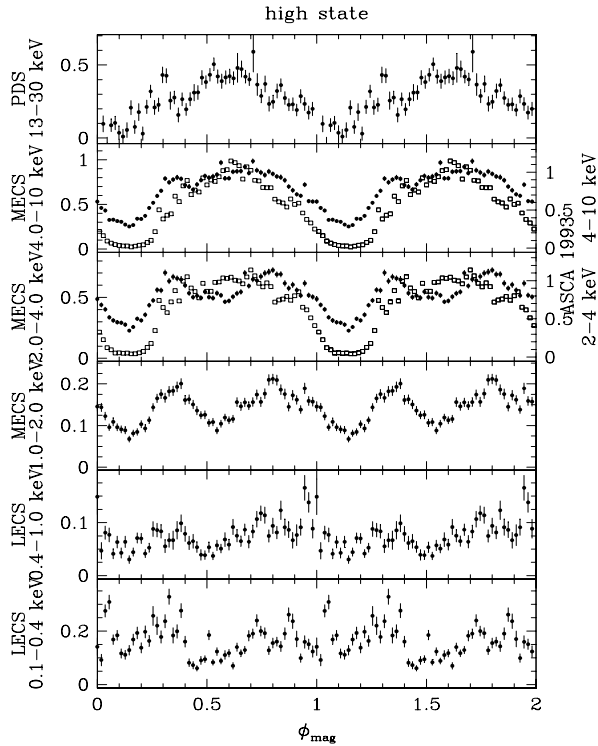, height=10.cm, width=8.cm}
\caption{ High state  LECS (0.1-0.4\,keV), (0.4-1\,keV), MECS (1-2\,keV),
(2-4\,keV), (4-10\,keV) and PDS (13-30\,keV) 
folded light curves (black dots) compared
to the same ASCA data (ordinates are scaled to the maximum). Flares
observed in the softer band have been removed before folding. Contrary to the
intermediate states of May and ASCA  these light curves reveal
a significant emission during the faint phase. 
A dip feature around $\phi \sim$ 0.45 (at high energies)
increasing in depth and moving towards
later phases at lower energies is also observed.}
\label{high_lc}
\end{figure}

In Fig.~\ref{high_lc} the folded light curves in different bands are  
shown together with the ASCA ones, all curves again
scaled to their respective maximum count rates.
The minimum is now significantly
above zero at 0.06\,$\rm cts\,s^{-1}$ in the soft (0.1-0.4\,keV),
at 0.257 \,$\rm cts\,s^{-1}$ in the 4--10\,keV band and at 
0.08 cts\,s$^{-1}$ in the 13--30\,keV band.
The BeppoSAX high state light curves are different 
from any previous X--ray dataset. Particularly
remarkable is the lack of orbital modulation in the 
soft bands ($\leq$ 1\,keV), although 
count rates are slightly lower at $\phi_{\rm mag} \sim$ 0.5
and $\sim$ 0.1. On the other hand, 
the hard ($\geq$ 2\,keV) X-ray light curves, though still keeping a 
sinusoidal-like shape
as in the intermediate state, have a well defined structured maximum,
with a dip occurring at 
at $\phi_{\rm mag} \sim$ 0.4 in the 13-30\,keV band, 
at $\sim 0.45$ in the 4--10\,keV range, and moving to later phases
at lower energies. 
At intermediate energies, in the 1--2\,keV range, this dip 
at $\phi_{\rm mag}\sim$0.55 almost reaches the same level as in the primary
minimum and mimics a double--humped light curve. The high count rate
at orbital minimum and the modification of the rise and decay shapes of
the maximum at high energies suggest a contribution from the secondary pole, 
which 
is expected to come into view between $\phi_{\rm mag} \sim $  0.9--0.4
(Heise et al. 1985). The lack of a clear soft X-ray modulation indicates that
AM\,Her was not in the so-called ``reverse" mode observed by EXOSAT in 1983 
(Heise et al. 1985),  where the secondary pole was dominating the soft X-ray
orbital modulation, but rather in an ``atypical" low soft X-ray mode, where
the soft X-ray emission from the main pole is comparable to 
that from the secondary  pole. An atypical, but not similar,  
behaviour was observed in AM\,Her during a high state in 1976 
(Priedhorsky et al. 1987) where no orbital modulation was detected 
in both soft and hard X-rays. At that epoch the source also showed a 
soft X-ray flaring activity. The soft X-ray flux was 
lower by a factor of three than that usually observed during high states. 
For this uncommon behaviour a shift of the
primary pole accretion column and corresponding changes in the absorption 
pattern were claimed (Priedhorsky et al. 1987). This interpretation 
however cannot work 
for the observed behaviour during the 1998 high state, since the 
lack of orbital modulation is detected only in the soft band. It is hence
more likely that the main accreting pole dominates the hard
X-ray emission while the secondary pole emission is essentially soft (see also
Sect. 4.2). A low soft X-ray emission at all phases could be produced by 
a decrease in the efficiency of blobby-accretion onto both primary and 
secondary poles (see also Sect. 5).

At intermediate energies a dip is clearly present in the light curve.
An inspection of hardness ratios in different bands reveals a clear
hardening at the maximum ($\phi_{\rm mag} \sim$ 0.55), suggesting 
absorption effects in the accretion column over the main accreting pole.

\begin{figure}
\epsfig{file=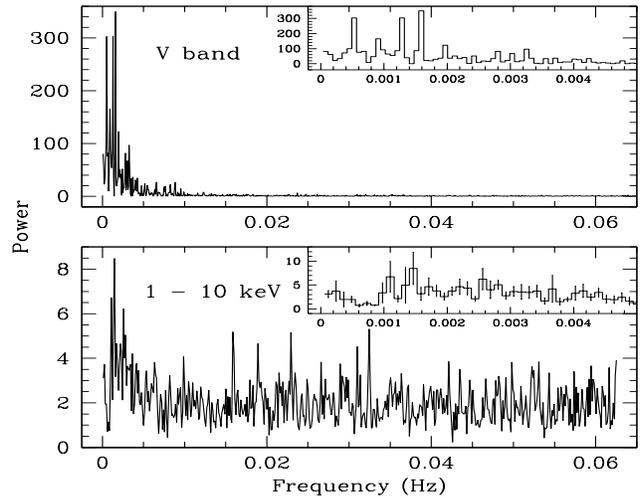, height=7.cm, width=8.8cm}
\caption{ The average power spectra in the high state of 5\,s resolution time
V filter data ({\it upper panel}) and of
8\,sec time resolution MECS data ({\it lower panel}). 
The normalization adopted is such that
the white noise level corresponds to a power of 2. The inserted enlargements
show the presence of peak features discussed in the text.}
\label{mecs_powspec}
\end{figure}

From Fig.~\ref{high_unfold_lc}, variations on time--scales shorter than 
the orbital period are apparent. MECS data have been therefore
inspected to search for short term variability. The MECS light curve
has then been detrended from long--term and 
orbital variations by subtracting a spline interpolation with a 202\,s
binning.  Then 16 intervals of continuous data with durations between 
$\sim$ 2200--3400\,s have been 
selected and for each of them the power spectrum has been computed. 
The resulting average spectrum of 8\,s resolution data, obtained with a
uniform rebinning of the individual spectra, is shown in the
lower panel of  Fig.~\ref{mecs_powspec}.
At high frequencies it goes rapidly to the noise level, corresponding
in Fig.~\ref{mecs_powspec} to a power of 2, while 
the low-frequency portion shows a rise at $\nu <$ 4\,mHz with a
shape similar to that inferred from  GINGA data 
by Beardmore \& Osborne (1997) and  interpreted as shot-noise. 
Peak features at 1.11\,mHz, 1.44\,mHz and 2.57\,mHz 
corresponding to 900\,s,  694\,s and 389\,s are also detected at $\sim 
3\,\sigma$ level, and possibly related to flickering produced
by inhomogeneities in the accretion flow.

\subsection {The optical variability}

Intermediate state white light photometry reveals  weak ($\sim$ 0.18\,mag)
orbital modulation but strong ($\ga$  0.4-0.5\,mag) flaring-type  
cycle-to-cycle variability which, together with the short duration of the
observations,  only allows one to identify a maximum 
around $\phi_{\rm mag} \sim 0.8$. The flare durations range between 
a few mins to a few tens of minutes.

\begin{figure}
\epsfig{file=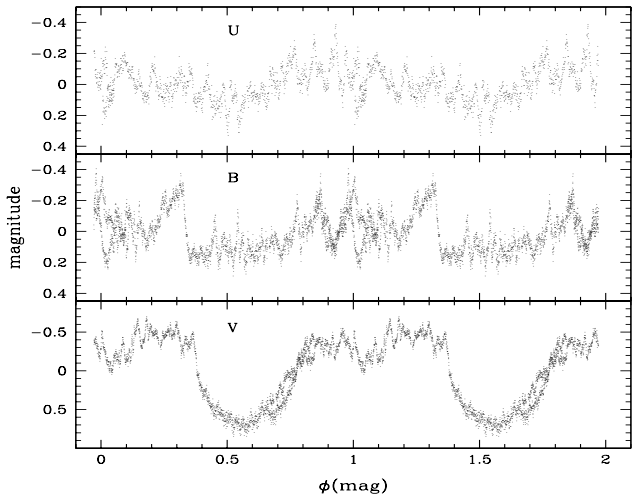, height=7.cm, width=8.8cm}
\caption{ High state U\,B\,V\, light curves of August 1998, binned
in 10\,s (U) and 5\,s  (B,V) intervals, showing
a marked difference between V band and U and B filters. A pronounced minimum 
at $\phi \sim 0.55 $ is a common feature, while a secondary minimum is 
clearly detected in the V light curve at $\phi \sim$ 0.1 but somewhat 
corrupted by short-term  flaring type variability in the other bands 
(see text). }
\label{amher_ubv_aug}
\end{figure}

Differently, the high state V band light curve shows a strong 
orbital behaviour, whilst U and B photometry have a less pronounced 
variability (Fig.~\ref{amher_ubv_aug}). In the V band the light curve is 
strongly non-sinusoidal with a well defined broad primary minimum ranging 
between $\phi_{\rm mag} \sim $ 0.4--0.7  and a  flat--topped maximum; 
a secondary minimum centered at $\phi_{\rm mag} \sim$ 0.1 is also apparent. 
Such shape of the V band modulation and the presence of a secondary minimum are
known to be characteristic of AM\,Her during high states (Mazeh et al. 1986
and reference therein;  Wickramasinghe et al. 1991; Beardmore \& Osborne 
1997). 
Exception is the peculiar behaviour observed in 1976, when AM\,Her was
lacking a clear orbital X-ray modulation and when the optical 
secondary minimum disappeared (Priedhorsky et al. 1987). 
On the other hand, the U and B band light curves are different,
though the primary minimum around $\phi_{\rm mag} \sim$ 0.5 is still observed
and the secondary minimum can  be possibly recognized in the B filter 
despite of the presence of short-term  variability (see below).
Noteworthy is that in the U band the shape is sinusoidal-like with a 
maximum around  $\phi_{\rm mag} \sim 0.9$. This is consistent with 
what observed in  May in white light and during other intermediate states 
(Beardmore \& Osborne 1997). However, there is a marked difference
with respect to low state optical observations (Bonnet-Bidaud et al. 2000),
where the light curves, especially in the U and B bands show similar 
phasing to that observed in the UV during low states,  
with a maximum at $\phi_{\rm mag} =0.6$, 
where the main accreting pole dominates (G\"ansicke et al. 1995).
Hence all this is in favour of a strong cyclotron beaming in the V band and
a substantial contribution from the secondary pole at shorter 
wavelengths. A detailed modeling of the V band orbital modulation with
cyclotron emission  is the subject of a forthcoming paper 
(G\"ansicke et al. in preparation).

\medskip

A marked flickering type variability is also observed in the optical data
with typical amplitudes of 0.1-0.5\,mag. The time--scale
 of this activity (a few minutes) is
different from that observed in the soft X-rays. In order to inspect 
these variations 
the long--term and orbital trends have been removed by subtracting a cubic 
spline interpolation of a large binning of the data. 
Data centered on the optical maximum and minimum have been selected,
and for each  of them the power spectrum has been computed. The average power
spectrum (upper panel of Fig.~\ref{mecs_powspec}) indeed shows not 
only a low frequency rise at $\nu \leq$ 2\,mHz 
but also peak features, at 0.5\,mHz, 1.3\,mHz and 1.6\,mHz, 
corresponding to 2000\,s, 769\,s and 625\,s at $\sim 5 \sigma$ level. 
These variations  are consistent with those inferred in hard X-rays, 
suggesting that they are related to the post shock accreting region.

\section{Spectral  analysis}

\subsection{Intermediate state}

\begin{figure}
\epsfig{file=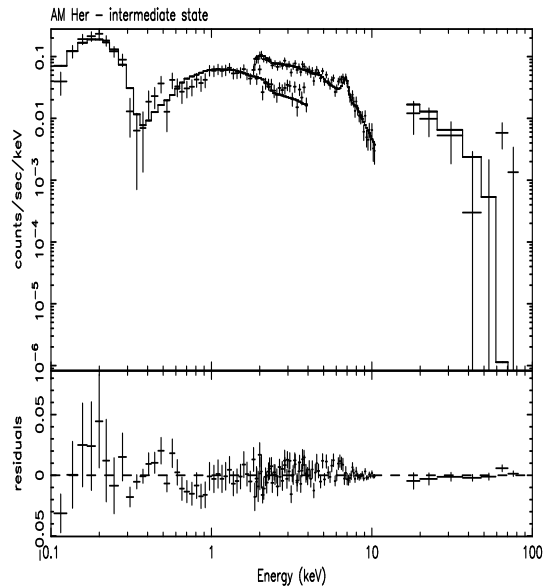, height=9.5cm, width=8.cm, angle=-90}
\caption{Data and residuals for the intermediate state, bright phase,
when fitted without partial absorption, the neutral iron line and the Compton
Reflection continuum.}
\label{int_badfit}
\end{figure}

The flux level of the source in this observation is not sufficient to allow
a detailed phase--resolved spectroscopy. The flux is virtually zero
at the minimum phase, and in order to increase  the signal--to--noise 
the spectrum has been selected over the bright phase between 
$\phi_{\rm mag}$=0.4 and 0.8. The results are
summarized in Table~\ref{fit_is}. 
(Hereinafter, the relative normalization between LECS and MECS has been left
free to vary to allow for the different time coverage, while that between
PDS and MECS has been kept fixed to 0.84). 
Firstly, we introduced only the very basic
spectral components for a polar: a black body, and an optically
thin thermal plasma emission (model {\sc mekal} in the spectral 
package XSPEC). The galactic
absorption has been fixed to the value of 9$\times10^{19}$ cm$^{-2}$
(G\"ansicke et al. 1995). The fit is poor 
($\chi^2_{\nu}$=1.37), and the iron 
abundance is quite large (2.8). Inspection of the residuals 
(Fig.~\ref{int_badfit}) suggests the presence of complex
absorption. The addition of a partial absorber
improves significantly  the quality of the fit; the plasma 
temperature diminishes considerably, and so does the iron abundance.
A further improvement in
the fit quality (and a decrease of both the temperature and the 
iron abundance) is achieved by adding a neutral iron line at 6.4 keV, 
as suggested by small residuals around that energy and by analogy
with previous GINGA and ASCA observations (Beardmore \& Osborne
1997; Ishida et al. 1997)   and with the high state data described below.  A
Compton reflection continuum (Matt 1999 and references therein), which 
is expected to be present along with the neutral iron line, has also been 
included, even if not required by the data. The improvement in the quality of
the fit is significant at the 97\% confidence level (F-test).
The value of $R$
in Table~2 represents the solid angle subtended by the cold matter in
units of 2$\pi$, assuming a viewing angle of 60$^{\circ}$

\subsection{High state}

\begin{figure}
\epsfig{file=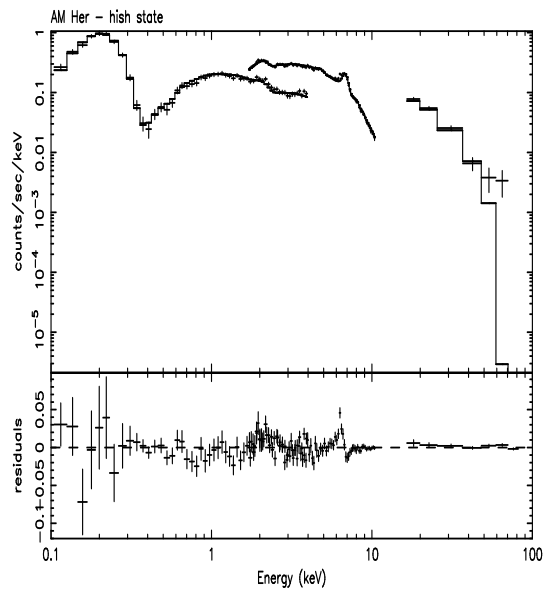, height=9.5cm, width=8.cm, angle=-90}
\caption{Data and residuals for the high state, bright phase,
when fitted without the neutral iron line and the Compton reflection
continuum.}
\label{noline}
\end{figure}

The much better statistics for the high state (due to both a higher flux and
a longer exposure time), together with the fact that in this state there is
significant emission also during the minimum, permits to perform a phase--resolved
spectroscopic analysis. The bright phase ($\phi$=0.4--0.9) spectrum has been 
constructed excluding the data during the flares discussed above.
The results are reported in Table~\ref{fit_hs},
where the same sequence of models have been applied. 
The simplest model (black body plus thermal plasma emission)
is now completely unacceptable. The
inclusion of partial absorption provides a dramatic improvement in the
fit, which however is still unacceptable. The neutral iron line
and the Compton reflection continuum are both required by the data 
(see Fig.~\ref{noline}). If only the line is included, the
$\chi^2$ is 144 for 127 d.o.f; apart from physical considerations, the
reflection continuum is required at the 99.7\% confidence level, according
to the F-test. 

Instead, a multi--temperature plasma and/or ionized or more complex
absorption do not improve the statistical quality of the fit. 

\begin{table*}
\centering
\caption{ Best fit parameters for the high state bright phase.}
\label{fit_hs}
\vspace{0.05in}
\begin{tabular}{lccccccccc}
\hline
\hline
~ & ~ & ~ & ~ & ~ & ~ & ~ & ~ & ~  \cr
\# & N$_{\rm H}$ &  C & $kT_{\rm bb}$ & $kT_{\rm h}$
& $A_Z$  & E.W.  & $R$ &  $\chi^2$/d.o.f. ($\chi^2_r$)   \cr
~ & (10$^{22}$ cm$^{-2}$) & ~ & (eV) & (keV) & ~ & (eV) & ~ & ~ \cr
~ & ~ & ~ & ~ & ~ & ~ & ~ & ~ & ~  \cr
\noalign {\hrule}
~ & ~ & ~ & ~ & ~ & ~ & ~ & ~ & ~  \cr
1 & ~ & ~ & 24.7$^{+1.1}_{-1.1}$ & 32.7$^{+2.0}_{-1.7}$ & 
5.09$^{+0.49}_{-0.44}$ & ~ & ~ & 1504/130 (11.6) \cr
2 & 4.97$^{+0.69}_{-0.59}$ & 0.51$^{+0.2}_{-0.2}$ & 24.7$^{+1.2}_{-1.1}$  
& 16.9$^{+0.9}_{-0.8}$ &    1.38$^{+0.15}_{-0.14}$  & ~ & ~ & 255/128 (1.99) \cr
3 & 3.56$^{+0.57}_{-0.49}$ & 0.50$^{+0.02}_{-0.02}$ & 25.0$^{+1.1}_{-1.2}$ & 
13.7$^{+2.2}_{-1.4}$ & 
0.89$^{+0.21}_{-0.14}$  & 107$^{+37}_{-26}$  & 1.43$^{+0.75}_{-0.74}$  & 134/126 (1.07) \cr
~ & ~ & ~ & ~ & ~ & ~ & ~ & ~ & ~  \cr
\hline
\hline
\end{tabular}
~\par
\end{table*}

\begin{figure}
\epsfig{file=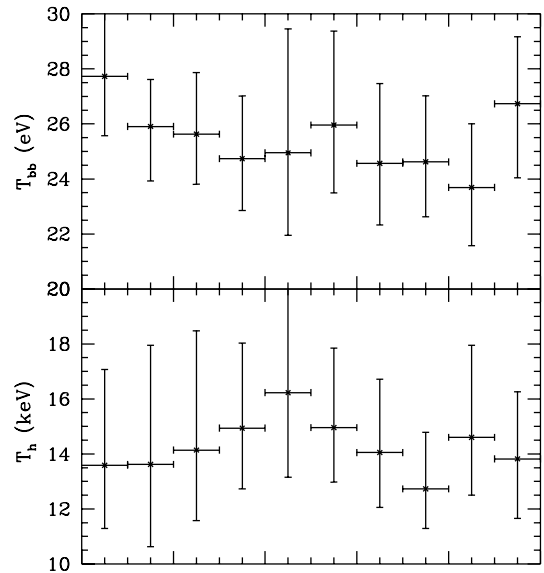, height=10.cm, width=10.cm}
\caption{Black body (upper panel) and thermal plasma (lower
panel) best fit temperatures as a function of the orbital phase.}
\label{temp}
\end{figure}

\begin{figure}
\epsfig{file=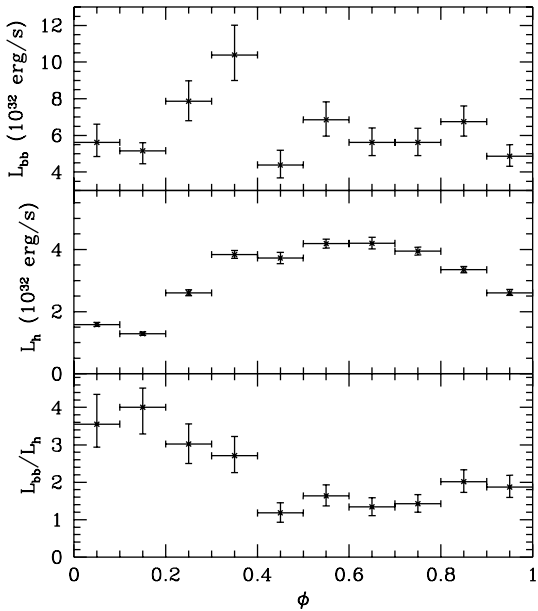, height=10.cm, width=10.cm}
\caption{Black body (upper panel) and thermal plasma (middle
panel) luminosities ($D$=91 pc), and their ratio (lower panel), as a function of the 
orbital phase.}
\label{leme_flux}
\end{figure}

The folded light curve has then been divided in 
10 phase bins, and the LECS+MECS 
spectra have been analysed for each bin. Again, flares were excluded.
The adopted model is the same as before, except for the 
reflection component which is an unnecessary complication at this level
of statistics. The iron abundance was fixed at the best fit value for
the bright phase, i.e. 0.89 times the solar value. In Fig.~\ref{temp}
the blackbody and plasma temperatures are shown as a function of phase.
Both temperatures are consistent with being constant.
Fixing them to their bright phase best
fit values (model 3 of Table~\ref{fit_hs}), we derived the bolometric 
luminosities of the two spectral components, which are shown in 
Fig.~\ref{leme_flux} along with their ratio. It is worth noting that 
the fluxes are corrected for absorption, and then any variation should be
intrinsic. At the
minimum the spectrum is significantly softer than at the maximum
(as also discussed in sect. \, 3.2 
where, however, effects of absorption were not separated). As the 
blackbody  temperature is constant, the increase
of the soft emission is due to an increase of the emitting area. The
simplest explanation is that the secondary pole is active. Furthermore, 
the fact that a similar
increase is not present in the hard component suggests that, at
the secondary pole, accretion mainly occurs in the form of 
high density blobs, shocking deep in the photosphere.

\begin{figure}
\epsfig{file=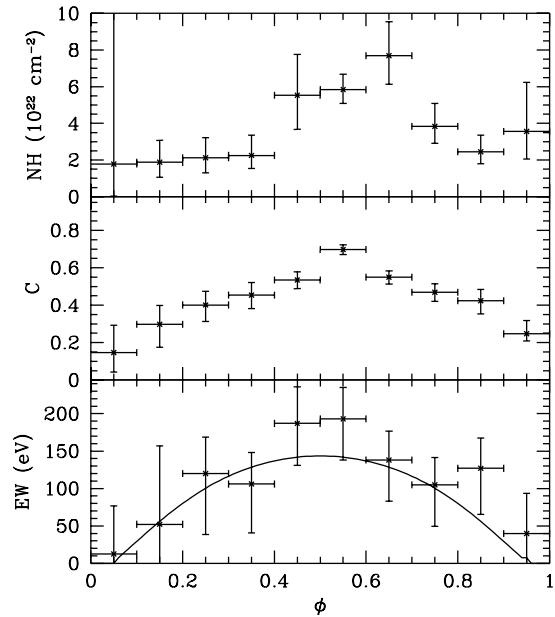, height=10.cm, width=10.cm}
\caption{Column density (upper panel) and covering factor (middle
panel) of the partial absorber, and the EW of the neutral iron line
(lower panel), as a function of the orbital phase. {\bf In the lower panel,
the expected values of the line EW are also shown (solid curve; see text 
for details).}}
\label{abs}
\end{figure}

A search for any phase--dependence of the partial absorption 
(Fig.~\ref{abs})  has also been performed.
The column density (upper panel) of the absorber has a
maximum around $ \phi_{\rm mag}$=0.4--0.7, while the covering 
factor (middle panel) 
increases steadily up to around $\phi_{\rm mag}$=0.5--0.6, and then decreases. 
This is not surprising, because at this phase the angle between the
line of sight and the accretion column is minimum, and then the 
amount of cold matter along the line of sight is likely to be the largest. 

In the lower panel of Fig.~\ref{abs}, the equivalent
width (EW) of the iron K$\alpha$ 6.4 keV line is plotted. 
The $\phi$--dependence of this
parameter is what expected if the line is originated in the white
dwarf photosphere illuminated by the hard X--ray radiation emitted by
the post--shock region. In fact, because the photosphere is optically
thick, the line EW has a well known angular behaviour (Matt et al. 
1991; George \& Fabian 1991; see also Matt 1999), decreasing with
the viewing angle. {\bf This is shown in the figure, where the expected
line EW (calculated following Basko 1978 and assuming: a plane parallel
geometry; a 13.7 keV bremsstrahlung illuminating spectrum; and an iron
abundance of 0.89 times the solar value, see Table~3) is also plotted 
(solid curve), adopting the 
inclination and colatitude of the primary pole as given by Cropper (1988). }

\medskip

Finally, the spectrum during flares has been analysed, summing 
data from all flares. The spectrum is significantly 
softer than during quiescence, whatever the phase. For an assumed distance of
91\,pc (G\"ansicke et al. 1995), the blackbody luminosity
is in fact 12.4$(^{+2.7}_{-1.9})$$\times10^{32}$ erg s$^{-1}$, while
that of the hard emission is
2.29$(^{+0.16}_{-0.12})$$\times10^{32}$ erg s$^{-1}$, the ratio being 
then $\sim$5.4. 
While the parameters of the hard emission are substantially the same,
the blackbody temperature is somewhat higher, 30.4$(^{+1.3}_{-1.1})\, \rm eV$.
Similar values for the flare and ``quiescence'' temperatures were found 
from ROSAT data  (Ramsay et al. 1996).

\section{Discussion and conclusions}

The BeppoSAX data presented here have shed new light onto the temporal
and spectral behaviour of AM\,Her.

The observations carried out at different brightness levels indicate
that the accretion mode can be substantially different. 
During intermediate states, accretion occurs mainly in the ``normal"
mode, with the primary pole dominating both hard and soft X-ray emissions.
A 20\,eV soft black body and a 14\,keV  temperature for the thermal plasma
are found, provided the inclusion of partial absorption, of 
a reflection component and of an iron 
$\rm K{\alpha}$ 6.4\,keV line with EW= 150\,eV.

High states instead can be very different from epoch to epoch. Differently
from previous observations, our BeppoSAX high  state observation 
shows a markedly different variability in the
soft and hard X-rays. A strong flaring activity characterized by 
an increase in flux by a factor of 5--10 with typical exponential
rise and decay time-scales of the order of tens of minutes is observed
below 0.4\,keV.  
In hard X-rays the variability is essentially related
to the orbital modulation. Differently from intermediate and previous high
state observations, the presence of significant emission during the 
faint phase in both soft and hard X-rays provide evidence that the 
secondary pole is emitting. Furthermore, the lack of an orbital 
variability in the soft X-ray ``quiescence'' emission contrasts with 
the strong modulation observed in hard  X-rays. This suggests that 
soft  X-ray emission from the primary pole 
(whatever produced by blobs or by reprocessing of hard X-rays) 
is highly inefficient. An atypical low
soft X-ray emission period was indeed observed in the past in AM\,Her during 
a high state  (Priedhorsky et al. 1987) but accompanied by the lack of 
the X-ray minimum in both soft and hard bands. While this was 
interpreted by a shift in the accretion column with a corresponding change of 
the accretion pattern, the behaviour observed in August 1998 appears to be
better explained by the onset of accretion onto the secondary pole whose
emission is mainly soft together with a less active soft X-ray production 
from the  primary pole.
 
It is worth noting that the soft flaring activity appears to be related to 
both accreting poles. The observed flaring activity is not uncommon for
AM\,Her  but with much shorter time scales. Furthermore similar variations 
are not observed  in the UBV coordinated photometry indicating that the soft 
X-ray flares are unrelated to the optical cyclotron post-shock emission.
Moreover, spectral analysis of flares indicates a higher temperature 
of the soft black body ($\sim$ 30\,eV) while the hard X-ray emission
has not varied.

The August high state BeppoSAX data 
have also revealed for the first time a strong energy dependence 
of the orbital modulation which moves from a single-humped shape at high
energies (4--30\,keV) to a double-humped shape around 1\,keV. This effect
can be explained in terms of absorption in the accretion column when pointing
towards the observer.


For this high state, it has been possible to perform a phase--resolved 
spectral analysis, and to measure the parameters 
of both black body and optically thin thermal plasma emission along the
orbital phase. The temperatures of both components are consistent with
being constant,  and
compatible with those inferred for the May intermediate state.
The plasma temperature (14 keV) also agrees with  the
ASCA data,  but not so the black body temperature, which is 25$\pm$1 eV 
(ASCA: 32$\pm$1 eV). The latter is however in agreement with EUVE (24.5 eV) 
(Ishida et al. 1997). The phase dependence of the 
partial absorber is indeed compatible with an increase of the amount of
cold matter when 
the accretion column is along the line of sight. The dip observed
at $\phi_{mag} \sim 0.5$ is consistent with such an 
increase, as both covering fraction and column density are higher.
The phase dependence of the EW of the neutral iron line
is consistent with an origin in the white dwarf surface 
illuminated by the hard X-rays from the post-shock regions.

\begin{figure}
\epsfig{file=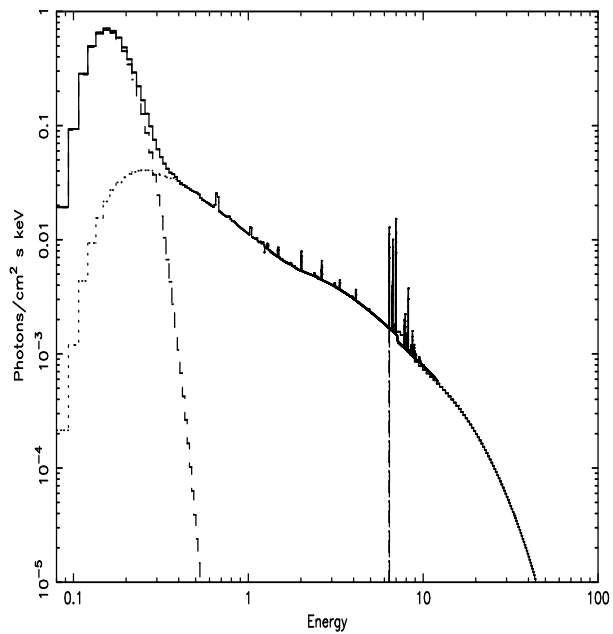, height=9.5cm, width=9.cm, angle=-90}
\caption{Best fit model for the high state, bright phase (model 3 in Table~3).
}
\label{model_high}
\end{figure}

Fixing the black body temperature of both intermediate and 
high states to 25\,eV
and that of the optically thin thermal plasma to 14\,keV, 
the bolometric luminosities of these components at the maximum, and adopting
$D$=91\,pc, are:
$\rm L_{BB} = 0.81(^{+0.31}_{-0.17}) \times 10^{32}\, erg\,s^{-1}$ and 
$\rm L_{Th.} = (0.683^{+0.027}_{-0.027}) \times  10^{32}\, erg\,s^{-1}$ for the
intermediate state; whilst for the high state these are 
$\rm L_{BB} = (5.32^{+0.31}_{-0.60}) \times 10^{32}\, erg\,s^{-1}$ and
$\rm L_{Th.} = (3.333^{+0.047}_{-0.037}) \times 10^{32}\, erg\,s^{-1}$.
We note that while the hard X-ray luminosities are typical of intermediate
and high states, that of the black body component during the high state
 is lower than previously measured. This is shown in Fig.~\ref{model_high}, 
which can be compared with that reported by Ishida et al. (1997) for the 
intermediate state in 1993.  The inferred 
soft-to-hard X-ray ratio is 1.2 and 1.6 during the intermediate and high 
state, respectively. 
It has been shown that in AM\,Her the bulk of reprocessed 
hard X-ray and cyclotron radiations emerges
in the UV range rather than in the soft X-rays (G\"ansicke et al. 1995), 
the latter being mainly produced by the blobby accretion mechanism. 
During the high state observed in August, both the inferred black body
luminosity and the lack of a clear modulation in the soft X-ray band suggest
that discrete accretion onto the primary pole is not as efficient as in 
other epochs. Given the low level of soft X-ray flux during the faint
phase it is not possible to infer the black body luminosity from the
secondary pole. 
From the high state black body luminosity  during the bright phase,
information on the mass accretion rate can be estimated. A value of 
$\rm 8.5 \times 10^{-11}\, M_{\odot}\,yr^{-1}$ for a white dwarf
mass of $\rm 0.6\,M_{\odot}$  is derived. A fractional accreting area  
$ f \sim  2.4 \times 10^{-4}$ is obtained, which is a factor of 
$\sim$ 3-10 larger than those derived by G\"ansicke et al. (1995) and
Ishida et al. (1997) respectively during typical high states. The local mass 
accretion rate then results to be $\sim$ 4\,g\,s$^{-1}$cm$^{-2}$ which is
too low to make blobby accretion efficient. 

As far as flares are concerned, the changes are essentially related to 
an increase in the temperature of the soft component, as also observed during 
other flaring epochs (Ramsay et al. 1996).  The increase in black body 
luminosity and then in the mass accretion rate is a factor of 2 with respect to 
``quiescence'', and comparable to usual high state values. 
Our estimates of the  fractional
area of the emitting black body during flares results to be similar to that 
during ``quiescence'' which gives a low local mass accretion rate 
of $\sim$ 8\,g\,s$^{-1}$cm$^{-2}$. However, since  we cannot exclude that 
accretion occurs onto a more localized area, we consider this value
as a lower limit.

Different from the flaring
variability are the rapid fluctuations detected in the hard
X-rays and optical ranges, the latter likely to be associated with 
inhomogeneities in the accretion flow itself. 

Furthermore, the optical orbital modulation observed in August is typical
of AM\,Her during high states, where the V band is dominated by cyclotron 
beaming from the primary pole (G\"ansicke et al. in preparation). 
The secondary pole instead appears to dominate at optical blue 
wavelengths. This is very different from low and other high states 
where the main accreting pole usually 
dominates the UV and the blue portion of the optical spectrum. 

\medskip

In summary, the present observations reveal that AM\,Her has switched from 
a ``normal" accretion mode during the intermediate state of May 1998 into a 
``two-pole" accretion mode during the high state of August 1998, with a less
efficient  ``blobby" accretion onto the primary pole.

\begin{acknowledgements}
We acknowledge the BeppoSAX SDC team (and in particular Paolo Giommi, 
Fabrizio Fiore and Angela Malizia) for providing pre--processed event files
and for their constant support in data reduction and analysis. 
We thank the BeppoSAX 
mission scientist, Luigi Piro, for his intelligent handling of the schedule,
and the mission planning team (Donatella Ricci, Milvia Capalbi and Sonia
Rebecchi) for the courtesy and patience.
The timely observations of the source in different states would not have been 
possible without the informations provided by AAVSO: we warmly
thank dr. J. Mattei and the entire AAVSO team.
We thanks Lucio Chiappetti for useful comments on an early version of the
manuscript. GM and DdM acknowledge financial support from ASI. 
\end{acknowledgements}

\end{document}